\documentclass[useAMS,usenatbib,usegraphicx]{mn2e}

\usepackage{amsmath}
\usepackage{amssymb}
\usepackage{times}

\title{On the Environment of Short Gamma--ray Bursts} 
\author[D. Kopa\v c, P. D'Avanzo, A. Melandri et al.]{D. Kopa\v c$^{1,2}$\thanks{E-mail: drejc.kopac@fmf.uni-lj.si}, P. D'Avanzo$^2$, A. Melandri$^2$, S. Campana$^2$, A. Gomboc$^{1,3}$, J. Japelj$^1$,
\newauthor M. G. Bernardini$^2$, S. Covino$^2$, S. D. Vergani$^2$, R. Salvaterra$^4$, G. Tagliaferri$^2$ \\
$^1$ University of Ljubljana, Faculty of Mathematics and Physics, Jadranska 19, SI--1000 Ljubljana, Slovenia\\
$^2$ INAF-Osservatorio Astronomico di Brera, via Bianchi 46, I--23807 Merate (LC), Italy \\
$^3$ Centre of Excellence SPACE-SI, A\v sker\v ceva cesta 12, SI--1000 Ljubljana, Slovenia \\
$^4$ INAF-IASF Milano, via E. Bassini 15, I-20133 Milano, Italy
}

\begin{document}



\maketitle

\begin{abstract}
In this paper we present a sample of $10$ short gamma--ray bursts (GRBs) with a robust redshift determination, discovered by the \textit{Swift} satellite up to January 2011. We measure their X-ray absorbing column densities and collect data on the host galaxy offsets. We find evidence for intrinsic absorption and no correlation between the intrinsic absorbing column density and the projected offset of the GRB from its host galaxy center. We find that the properties in the gamma regime ($\mathrm{T}_{90}$, fluence and 1-s peak photon flux) of short GRBs with ``bright'' and ``faint'' X-ray afterglow likely disfavour different prompt emission mechanisms. Host galaxy offset and GRB duration ($\mathrm{T}_{90}$) do not correlate. Instead, there is a hint of anti-correlation between the effective radius normalised host galaxy offset and $\mathrm{T}_{90}$. Finally, we examine the properties of short GRBs with short-lived and long-lived X-ray afterglows, finding that some short GRBs with short-lived X-ray afterglows have their optical afterglow detected. In light of this, the X-ray afterglow duration does not seem to be an unique indicator of a specific progenitor and/or environment for short GRBs.
\end{abstract}


\section{Introduction}
\label{sect:intro}

\par Gamma--ray bursts (GRBs) are cosmic explosions that release an extreme amount of energy in a very short time. As it was already noticed in the 1980s \citep[e.g.,][]{Norris1984} and became more obvious later, GRBs form two distinct populations \citep[e.g.,][]{Kouveliotou1993}: the short and long GRBs (SGRBs and LGRBs, respectively), defined at first approximation on the basis of the burst duration (SGRBs lasting less than $\sim 2\,\mathrm{s}$ in the observer frame), and likely corresponding to two different progenitors.

\par The merger of a double neutron star (NS-NS) or a neutron star -- black hole (NS-BH) binary system is currently the leading model for SGRBs. The events predicted by this model \citep[e.g.,][]{Eichler1989, Narayan1992, Nakar2007} are expected to have comparable time scale and energy release to those observed in SGRBs. In such systems, the delay between binary formation and merging is driven by the gravitational wave inspiral time, which is strongly dependent on the initial system separation. Some systems are thus expected to drift away from the star-forming regions in which they formed, before merging takes place. Simulations \citep{Belczynski2002, Belczynski2006} show that a large fraction of the merging events should take place in the outskirts or even outside the galaxies, in low density environments. A much faster evolutionary channel has been proposed \citep{BelczynskiKalogera2001, PernaBelczynski2002, Belczynski2006}, leading to merging in only $\sim 10^6-10^7\,\mathrm{yr}$, when most systems are still immersed in their star-forming regions. The above scenarios are based on ``primordial'' binaries, i.e., systems that were born as binaries. Alternatively, a sizeable fraction of NS-NS systems may form dynamically by binary exchange interactions in globular clusters during their core collapse \citep{Grindlay2006, Salvaterra2008}. The resulting time delay between star formation and merging would be dominated by the cluster core-collapse time and thus be comparable to the Hubble time \citep{Hopman2006}.

\par Merger scenario differs from the collapse of a massive star, which is believed to be associated with long duration GRBs (LGRBs). These two types of progenitors produce different outcomes when exploding as GRBs. At variance with LGRBs, we have no ``smoking gun'' (like supernova signatures) to identify the nature of the progenitors of SGRBs. However, SGRBs are not distinguished from LGRBs only by their duration, but also by other observed properties. If we consider the prompt emission, negligible spectral lag \citep{Norris2000,Norris2001} and hard spectra \citep{Kouveliotou1993} are common for SGRBs. As opposed to LGRBs, for which the isotropic equivalent gamma-ray energy, $E_{\gamma,\mathrm{iso}}$, is of the order of $10^{53}\,\mathrm{erg}$ and for which the host galaxies are typically dwarf galaxies with high star formation rate \citep{Fruchter2006, Savaglio2009}, SGRBs are typically less energetic ($E_{\gamma,\mathrm{iso}}$ is of the order of $10^{49} - 10^{51}\,\mathrm{erg}$), they occur in both early- and late-type galaxies with lower star formation rate and are associated with an old stellar population \citep{Nakar2007, Berger2009, Berger2011}. Furthermore, SGRBs have been found to be inconsistent with the $E_\mathrm{p,i}-E_\mathrm{iso}$ correlation (\citealt{Amati2006, Amati2007, Amati2008}).

\par The afterglows of SGRBs tend to be significantly fainter on average than those of LGRBs \citep{Kann2011}. This is believed to be a consequence of the energetics and the surrounding environment \citep{Nakar2007}. As shown in \citet{Campana2010}, a powerful tool to characterise the GRB environment is the study of their X-ray absorbing column densities. By a systematic analysis of LGRBs with known redshift promptly observed by the \textit{Swift} X-Ray Telescope (XRT), \citet{Campana2010} found clear evidence that LGRB X-ray afterglows are heavily absorbed and occur in dense environments, as expected in the context of a massive stellar progenitor. 

\par In this paper we present a comprehensive analysis of the full sample of SGRBs with robust redshift determination, promptly observed\footnote{Within $150\,\mathrm{s}$ from the burst occurrence, and without the autonomous slew delay due to an observing constraint or due to a low merit value.} by the \textit{Swift} XRT up to January 2011. For all these events we derived the intrinsic X-ray column densities. Our findings are then compared to the results of \citet{Campana2010} obtained for LGRBs, with the aim of checking if the surrounding environment of these two classes of events is different and if this is perhaps related to the type of progenitor (as already discussed by \citealt{Salvaterra2010}), as well as to various SGRBs properties (redshift, duration, host galaxy offset and normalised host galaxy offset).

\par In Section \ref{sect:data}, we present the analysis of X-ray data taken from the \textit{Swift} XRT and describe how our sample was built. In Section \ref{sect:results}, we perform various analyses on our sample and discuss the results. Summary and conclusions are given in Section \ref{sect:conclusion}. 

\par Throughout the paper we assume a standard cosmology with parameters: $H_0 = 71\,\mathrm{km\,s^{-1}\,Mpc^{-1}}$, $\Omega _\Lambda = 0.73$, $\Omega _\mathrm{M} = 0.27$.

\section{Data Analysis}
\label{sect:data}

\subsection{Sample selection}
\label{sect:sample_selection}
\par We collected the information on the $\mathrm{T}_{90}$\footnote{Time interval in which $90\%$ of the fluence in gamma-rays (in this case, in the $15-350\,\mathrm{keV}$ energy band) is detected.} in the observer's frame from the \textit{Swift}-BAT refined analysis GCN circulars\footnote{http://gcn.gsfc.nasa.gov/gcn3\_archive.html.}, together with the properties in gamma regime (spectral hardness and spectral lag) for all GRBs detected with the \textit{Swift} Burst Alert Telescope (BAT) until January 2011. We consider all GRBs classified as short in the Swift-BAT refined analysis GCN circulars$^3$, where additional properties in the gamma regime (apart from $\mathrm{T}_{90}$), such as the lack of a spectral lag and the hardness ratio are used to assess the short nature of a GRB. These criteria enable to include in our sample also SGRBs with an extended emission (EE), for which $\mathrm{T}_{90}$ can be well above $2\,\mathrm{s}$. We selected the events observed by the \textit{Swift} XRT and obtained a list of $60$ SGRBs. 

\par To exclude any observational biases, we checked the time delay between the BAT trigger and the start of the observations by the XRT. We found that for $13$ SGRBs the XRT observations started with significant delay of hours or even days after the BAT trigger due to observing constrains. We eliminated these SGRBs and ended up with $47$ SGRBs. Among these, $6$ had no X-ray afterglow detected and $17$ had an X-ray afterglow too faint to perform any spectral analysis (see Section \ref{subsect:analysis}). Also these $23$ SGRBs have been excluded from our analyses.

\par For the remaining $24$ SGRBs (see Table \ref{tab:he_properties} for the complete list) we retrieved the redshift information from the literature. We used only redshifts that are robust, meaning that an optical afterglow (OA) was detected and found to lie within the host galaxy's light with a sub-arcsecond precision (so that the association between a host galaxy and a GRB is clear), and that the spectrum of the associated host galaxy was recorded. We obtained robust redshifts for $13$ SGRBs, mainly from the GCN Circulars Archive or from published papers. We put the remaining $11$ SGRBs to redshift $z=0$, to extend our analysis on SGRBs' intrinsic X-ray absorption with the obtained lower limits.

\par The sample of $13$ SGRBs with robust redshifts includes also $3$ GRBs for which the classification as a SGRB is still debated; GRB~060614 was a supernova-less GRB at $z=0.125$ down to very deep optical limits, with spectral lag typical for SGRBs, but with a $\mathrm{T}_{90}$ of $102\,\mathrm{s}$ and time-averaged spectral properties similar to LGRBs (\citealt{Gehrels2006, DellaValle2006, Mangano2007, Amati2007}). The possibility that this event is a SGRB with an EE was also suggested (\citealt{Fynbo2006, GalYam2006, Zhang2007}). GRB~090426 can be classified as a SGRB based on its $\mathrm{T}_{90}$, but the spectral and energy properties are more similar to those of LGRBs (\citealt{Levesque2009, Antonelli2009, Ukwatta2009, Thone2011}). Similarly, GRB~100724A can be classified as a SGRB based on its $\mathrm{T}_{90}$, but spectral lag and hardness of the spectrum point towards a LGRB classification \citep{Ukwatta2010}. Given the uncertainty in the classification, we do not consider also these $3$ GRBs in our analyses and thus end up with $10$ SGRBs with robust redshift determination. 

\par Another selection criterion for SGRBs could be the inconsistency with the $E_\mathrm{p,i}-E_\mathrm{iso}$ correlation (\citealt{Amati2007}), in which case one additional GRB could be considered short: GRB~060505 (see also \citealt{Ofek2007, Thone2008, Mcbreen2008, Xu2009}). However, GRB~060505 was not observed promptly by the \textit{Swift} XRT and so we did not include it in our analyses.

\par For the sake of simplicity throughout the paper we name as Sample \MakeUppercase{\romannumeral 1} the sample of $10$ SGRBs with robust redshift determination (presented in Table \ref{tab:sample}), and as Sample \MakeUppercase{\romannumeral 2} the sample of $11$ SGRBs without robust redshift determination (presented in Table \ref{tab:sample_lowerlimits}).

\begin{table*}
\renewcommand{\arraystretch}{1.1}
\begin{center}
\tabcolsep 3pt
\begin{tabular}{ccccccccccccc}
\hline
& 

& 
& 
& 
\multicolumn{6}{c}{X-ray afterglow properties} & 
\multicolumn{2}{c}{Host galaxy} & 
Ref. \\
GRB & 
$z$ & 
$\mathrm{T}_{90}$ &
EE & 
SL/LL & 
$\mathrm{N_H}$ Gal. & 
Exp. (interval) &
$\Gamma$ &
$\mathrm{N_H}(z)$ & 
C-stat (dof) & 
Offset &
Norm. offset &
\\
& 
&
[s] &  
& 
& 
[$10^{20}\,\mathrm{cm^{-2}}$] &
[$\mathrm{ks}$] ([$\mathrm{s}$]) &
&
[$10^{21}\,\mathrm{cm^{-2}}$] &
&
[$\mathrm{kpc}$] &
&
\\
\hline
050724 & $0.257$ & $3.0$ & yes & LL & $14.0$ & $22.0$ ($6000 - 10^5$) & $1.66_{-0.20}^{+0.15}$ & $2.1_{-1.3}^{+1.5}$ & $214.5 (279)$ & $2.54 \pm 0.08$ & $0.43 \pm 0.02$ & a) \\
051221A & $0.546$ & $1.4$ & no & LL & $5.7$ & $69.1$ ($6000-2 {\times} 10^5$) & $2.10_{-0.10}^{+0.14}$ & $2.2_{-1.1}^{+1.1}$ & $260.4 (307)$ & $1.53 \pm 0.31$ & $0.30 \pm 0.06$ & a) \\
061006 & $0.438$ & $130.0$ & yes & LL & $14.1$ & $20.6$ ($150-10^5$) & $1.84_{-0.27}^{+0.28}$ & $< 2.6^\mathrm{UL}$ & $104.1 (170)$ & $1.44 \pm 0.29$ & $0.40 \pm 0.08$ & a) \\
070714B & $0.923$ & $64.0$ & yes & LL & $6.4$ & $25.9$ ($450-6 {\times} 10^4$) & $1.97_{-0.16}^{+0.11}$ & $2.8_{-2.1}^{+2.1}$ & $283.6 (292)$ & $3.08 \pm 0.47$ & $0.78 \pm 0.12$ & a) \\
070724A & $0.457$ & $0.4$ & no & LL & $1.2$ & $0.024$ ($82-106$) & $1.51_{-0.23}^{+0.24}$ & $5.5_{-2.5}^{+2.8}$ & $161.9 (232)$ & $4.76 \pm 0.06$ & $1.12 \pm 0.02$ & a) \\
071227 & $0.381$ & $1.8$ & yes & LL & $1.3$ & $0.05$ ($90-140$) & $1.51_{-0.17}^{+0.17}$ & $2.8_{-1.1}^{+1.3}$ & $247.8 (330)$ & $14.8 \pm 0.3$ & $1.10 \pm 0.03$ & b) \\
080905A & $0.122$ & $1.0$ & no & / & $9.0$ & $0.9$ ($330-1230$) & $1.75_{-0.52}^{+0.56}$ & $< 3.4^\mathrm{UL}$ & $94.2 (76)$ & $18.1 \pm 0.4$ & / & a) \\
090510 & $0.903$ & $0.3$ & no & LL & $1.7$ & $0.14$ ($100-240$) & $1.79_{-0.15}^{+0.16}$ & $1.9_{-1.4}^{+1.5}$ & $284.1 (315)$ & $7.8 \pm 3.9$ & $1.29 \pm 0.65$ & c) \\
100117A & $0.915$ & $0.3$ & no & SL & $2.7$ & $0.05$ ($105-155$) & $1.44_{-0.22}^{+0.25}$ & $3.4_{-3.0}^{+4.4}$ & $156.6 (215)$ & $0.47 \pm 0.31$ & / & d) \\
100816A & $0.805$ & $2.9$ & yes & LL & $4.5$ & $16.1$ ($200-5 {\times} 10^4$) & $1.90_{-0.15}^{+0.13}$ & $2.3_{-1.5}^{+1.6}$ & $273.7 (309)$ & $8.2 \pm 2.3$ & $0.64 \pm 0.20$ & c) \\
\hline
\end{tabular}
\end{center}
\caption{Properties of $10$ SGRBs with robust redshifts (Sample \MakeUppercase{\romannumeral 1}). Columns are: GRB identifier, redshift, $\mathrm{T}_{90}$, SGRB with an extended emission (EE), short-lived (SL) or long-lived (LL) X-ray afterglow (see Section \ref{sl_ll}), Galactic X-ray absorbing column density, X-ray spectrum exposure time and time interval, photon index ($\Gamma$), intrinsic X-ray absorbing column density, Cash's C statistic of the spectral fit (with the corresponding degrees of freedom), host galaxy offset, normalised host galaxy offset, references. All SGRBs have an optical afterglow (OA) detected. Redshifts are obtained from \citet{Berger2009} and references therein, except for GRB~071227 \citep{DAvanzo2009}, GRB~080905A \citep{Rowlinson2010a}, GRB~090510 \citep{Mcbreen2010}, GRB~100117A \citep{Fong2011} and GRB~100816A \citep{Tanvir2010}. Errors are given at $90\%$ confidence level. $^\mathrm{UL}$ indicates upper limits. For GRB~070724A, GRB~071227, GRB~090510 and GRB~100117A we used WT mode spectra, while for others we used PC mode spectra in the specified time interval. For GRB~080905A we can not determine if it is SL or LL, due to the lack of the X-ray flux information around $10^4\,\mathrm{s}$ after the trigger. \newline References for host galaxy offsets: a) \citet{Church2011} and references therein; b) \citet{Fong2010, Berger2011}; c) Host galaxy offset from our inspection; d) \citet{Fong2011}.
} 
\label{tab:sample}
\end{table*}

\subsection{Sample analysis}
\label{subsect:analysis}
\par In order to determine intrinsic absorption in X-rays for SGRBs, we performed a similar analysis to the one presented by \citet{Campana2010} for the LGRBs. We used the \textit{Swift} XRT GRB lightcurve repository \citep{Evans2009}. Because SGRBs have significantly fainter afterglows as opposed to LGRBs \citep{Kann2011}, it is sometimes impossible to get enough X-ray photons to perform any spectral analysis\footnote{We required at least $200$ counts with constant $0.3-1.5\,\mathrm{keV}$/$1.5-10\,\mathrm{keV}$ hardness ratio in the time interval that is specified in Table \ref{tab:sample} and Table \ref{tab:sample_lowerlimits} to perform spectral analysis and obtain the values of the intrinsic X-ray absorbing column density.}. In addition, typical X-ray afterglow light curves have a multi-component canonical shape and show strong variability and spectral evolution especially at early times. We try to avoid taking data from this epoch and therefore we loose the brightest part of the afterglow. Based on these facts, we used data mostly from photon counting (PC) mode, except for very dim afterglows, where we were forced to use also data from window timing (WT) mode to increase our statistics by gathering enough X-ray photons. We analysed data only in time epoch where the $0.3-1.5\,\mathrm{keV}$ to $1.5-10\,\mathrm{keV}$ hardness ratio is constant, in order to prevent spectral changes that can affect the X-ray column density determination.

\par We used the \textit{Swift} XRT GRB spectrum repository\footnote{http://www.swift.ac.uk/xrt\_spectra/.} to obtain the intrinsic X-ray absorbing column densities. We also checked that the spectral fits are consistent with the values reported on the repository. The spectra that we used were binned to have at least one count in each spectral bin, so that the C-statistic can be used for fitting. Using \texttt{XSPECv12.6} software \citep{Arnaud1996} we fitted the spectra with the combination of power-law behaviour and photoelectric absorption contributions (phabs$^\ast$zphabs$^\ast$powerlaw) in the $0.3-10\,\mathrm{keV}$ energy range. We fixed one contribution of photoelectric absorption to the value of the Galactic equivalent X-ray column density along the direction of a GRB using \citet{Kalberla2005}. Since there is an error associated with these values \citep{Wakker2011}, we added an uncertainty factor of $0.1 \times (1 + z)^{2.6}$ (i.e., $10\%$, propagated with the redshift) to the intrinsic error \citep{Watson2011}. The second contribution, shifted in energy to the rest-frame of the GRB using its redshift (for SGRBs from Sample \MakeUppercase{\romannumeral 2} we used $z=0$), was left as a free parameter to vary. The intrinsic X-ray absorbing column densities for our final sample are provided in Table \ref{tab:sample}.

\par Working on a sample of $10$ SGRBs observed with the Hubble Space Telescope, \citet{Fong2010} showed that while the host galaxy offsets (offset of a projected GRB optical afterglow location from the center of its host galaxy) of SGRBs are larger than those of LGRBs, the distribution of normalised host galaxy offsets (offset divided by the effective radius $R_e$ of the galaxy) is nearly identical due to the larger size of SGRBs' host galaxies. For each SGRB belonging to Sample \MakeUppercase{\romannumeral 1} we compared the derived X-ray column density with the value of the projected host galaxy offset and with the normalised host galaxy offset. Some SGRBs from Sample \MakeUppercase{\romannumeral 1} have their host galaxy and normalised host galaxy offset already presented in the sample of \citet{Fong2010}. However, for the sake of homogeneity, we computed independently the host galaxy effective radii for the SGRBs of Sample \MakeUppercase{\romannumeral 1}, using public archival and proprietary ground-based imaging data. All SGRBs' host galaxies were observed with the ESO-VLT, equipped with the FORS1/2 camera, with the exception of GRB~070714B (Gemini-N/GMOS data) and GRB~100816A (TNG/DOLORES data). Using the extended surface photometry tool of the GAIA\footnote{Graphical Astronomy and Image Analysis Tool, http://astro.dur.ac.uk/$\sim$pdraper/gaia/gaia.html} package, we obtained the surface brightness profiles of $8$ SGRBs' host galaxies. The obtained galaxy profiles were fitted with a S\'ersic model:
\begin{equation}
\label{sersic}
I(R) = I_e \, \mathrm{e}^{-b_n[\left({\frac{R}{R_e}}\right)^{1/n}-1]} \,,
\end{equation}
where $n$ is the index indicating the profile curvature ($n=1$ gives an exponential disk profile, $n=4$ is the de Vaucouleurs profile), $b_n$ is a dimensionless scale factor that can be approximated with $b_n~\sim~2n~-~\frac{1}{3}~+~\frac{4}{405n}~+~\frac{46}{25515n^2}$ \citep{Ciotti1999}, $R_e$ is the galaxy effective radius and $I_e$ is the intensity at $R=R_e$. For each fit, $n$, $R_e$ and $I_e$ were left free to vary. Host galaxy offsets and normalised host galaxy offsets for SGRBs of Sample \MakeUppercase{\romannumeral 1} are provided in Table \ref{tab:sample}.

\par To check for various correlations we used the Spearman's rank correlation test throughout the paper. The uncertainties in the correlation coefficients were estimated with a simple Monte Carlo simulation, by taking into account the true errors using
\begin{equation}
\label{eq:mc}
\mathrm{value^{simulation}} = \mathrm{value^{true}} + \epsilon \times \mathrm{error^{true}}\,,
\end{equation}
where $\epsilon$ is a random number drawn from a normal distribution with zero mean and unit variance. With this method we obtained $1000$ simulated correlation coefficients for each set of values, and estimated the uncertainty on the original correlation coefficient as the standard deviation of simulated coefficients. Together with the correlation coefficient and its uncertainty, the $p$-value is also given.

\section{Results and Discussion}
\label{sect:results}
\par We were primarily interested in the study of the intrinsic X-ray absorption, and its correlations with $\mathrm{T}_{90}$, redshift, and particularly with host galaxy offset and normalised host galaxy offset. We extended our analysis with the discussion on the properties in the gamma energy band, on the SGRBs with an extended emission and on the short-lived and long-lived X-ray afterglows (defined in light of the X-ray flux measured at $10^4\,\mathrm{s}$ after the burst; see Section \ref{sl_ll}). The study of these observables, which could be commonly obtained for SGRBs, provides the answer whether properties at high energies (such as $\mathrm{T}_{90}$, energetics and X-ray afterglow duration) are linked to the properties of the environment.

\subsection{Intrinsic X-ray absorption}
\label{sect:intrinsic_absorption}
\par
\begin{figure}
\begin{center}
\includegraphics[angle=-90,width=1\columnwidth]{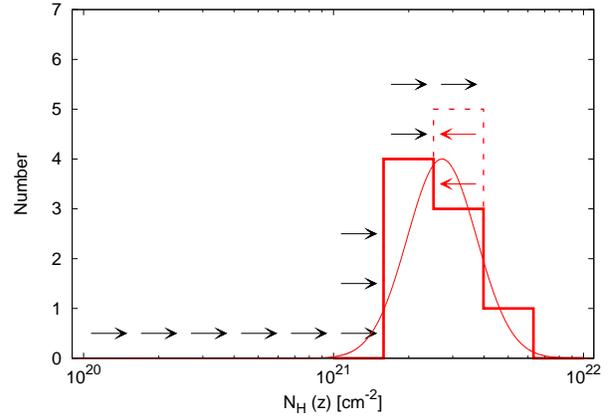}
\end{center}
\caption{Intrinsic X-ray absorbing column densities ($\mathrm{N_H}(z)$) distribution for Sample \MakeUppercase{\romannumeral 1} (red histogram). Red dashed line and arrows represent $2$ SGRBs from Sample \MakeUppercase{\romannumeral 1} for which only upper limits were determined. Black arrows represent the lower limits of the $\mathrm{N_H}(z=0)$ value for $11$ SGRBs from Sample \MakeUppercase{\romannumeral 2} (we put them to $z=0$). Continuous red solid line is the log Gaussian fit for $8$ SGRBs from Sample \MakeUppercase{\romannumeral 1} without upper limits.}
\label{fig:nhdist}
\end{figure}

 We first derived the intrinsic X-ray absorbing column densities, $\mathrm{N_H}(z)$, for SGRBs belonging to Sample \MakeUppercase{\romannumeral 1}. We obtained a measure for $8$ of them, while for $2$ we obtained only upper limits (Table \ref{tab:sample}). Figure \ref{fig:nhdist} shows the distribution of $\mathrm{N_H}(z)$ for our whole sample. The distribution for the $8$ SGRBs with direct estimates of the intrinsic column density can be well fitted by a log Gaussian function, with a mean $\mathrm{log\,N_H} (z) = 21.4$ and a standard deviation $\sigma _{\mathrm{log\,N_H}(z)} = 0.1$ (reduced $\chi ^2$ of the fit is $0.4$ with $2$ d.o.f.). This shows that SGRBs are intrinsically absorbed.

\par Nevertheless, the mean $\mathrm{log\,N_H} (z)$ for SGRBs is on average lower than for LGRBs from \citet{Campana2010}, who reported a value of $21.9$ with a standard deviation of $0.5$. However, given the quoted error bars, the values are consistent. But it is worth to note that SGRBs from Sample \MakeUppercase{\romannumeral 1} span in redshift up to $z_\mathrm{max}^\mathrm{SGRBs} = 0.923$, so we are biased to lower redshifts when comparing $\mathrm{N_H}(z)$ values, since LGRBs from \citet{Campana2010} span in redshift up to $z_\mathrm{max}^\mathrm{LGRBs}=8.1$. To make a more appropriate comparison, we determined a mean $\mathrm{log\,N_H} (z)$ for LGRBs with $z < 0.923$ and $z > 0.923$ in the sample from \citet{Campana2010}, obtaining a mean $\mathrm{log\,N_H} (z<0.923) = 21.5$ with a standard deviation of $0.4$ (reduced $\chi ^2$ of the fit is $1.5$) and a mean $\mathrm{log\,N_H} (z>0.923) = 22.0$ with a standard deviation of $0.4$ (reduced $\chi ^2$ of the fit is $0.6$), respectively. Thus, by limiting the analysis to $z<0.923$, the mean $\mathrm{N_H}(z)$ of the LGRBs and the SGRBs from Sample \MakeUppercase{\romannumeral 1} do not differ significantly.

\par To further investigate if the $\mathrm{N_H}(z)$ distributions of SGRBs and LGRBs are different, we used the two-sample Kolmogorov-Smirnov (K-S) test. We tested the distributions of $8$ SGRBs from Sample \MakeUppercase{\romannumeral 1} and $85$ LGRBs from \citet{Campana2010}, obtaining the value of the K-S statistic $D_1 = 0.70$, with the corresponding probability that the two distributions are not drawn from different populations $P_1 = 7 \times 10^{-4}$. Using the subsample of $14$ LGRBs with $z<0.923$ instead, we obtained $D_2 = 0.52$ and $P_2 = 0.1$, indicating that the two distributions are likely drawn from the same population. After the submission of this work, \citet{Margutti2012} independently reached similar conclusions on the intrinsic X-ray column density distribution of SGRBs.

\begin{table}
\begin{center}
\tabcolsep 2.3pt
\begin{tabular}{cccccc}
\hline
GRB & 
$\mathrm{T}_{90}$ &
OA & 
SL/LL & 
Exp. (interval) &
$\mathrm{N_H}(z=0)$
\\
& 
[s] &  
&
& 
[$\mathrm{ks}$] ([$\mathrm{s}$]) &
[$10^{21}\,\mathrm{cm^{-2}}$] 
\\
\hline
051227$^\mathrm{EE}$ & $8.0$ & yes & LL & $0.05$ ($115 - 160$) & $> 1.3$\\
060313 & $0.7$ & yes & LL & $22.7$ ($4000 - 7{\times} 10^4$) & $> 0.1 $\\
060801 & $0.5$ & no & SL & $13.3$ ($50 - 4{\times}10^4$) & $> 0.3$\\
080503$^\mathrm{EE}$ & $170.0$ & yes & SL & $0.06$ ($90 - 150$) & $> 0.5$\\
080702A & $0.5$ & no & SL & $8.4$ ($50 - 2{\times}10^4$) & $>1.7$ \\
080919 & $0.6$ & no & SL & $1.9$ ($100 - 2000$) & $> 1.7$\\
081226A & $0.4$ & no & SL & $13.9$ ($100 - 5{\times} 10^4$) & $> 3.5$\\
090515 & $0.04$ & yes & SL & $0.06$ ($80 - 140$) & $> 0.2$\\
090607 & $2.3$ & no & SL & $0.03$ ($103 - 133$) & $> 1.0$\\
100702A & $0.16$ & no & SL & $15.0$ ($250 - 5{\times} 10^4$) & $> 1.3$\\
101219A & $0.6$ & no & / & $0.85$ ($50 - 900$) & $> 0.9$\\
\hline
\end{tabular}
\end{center}
\caption{Properties and lower limits of the $\mathrm{N_H}$ values at $z=0$ for $11$ SGRBs without robust redshifts (Sample \MakeUppercase{\romannumeral 2}). Columns are: GRB identifier, $\mathrm{T}_{90}$, optical afterglow (OA) detection, short-lived (SL) or long-lived (LL) X-ray afterglow (see Section \ref{sl_ll}), X-ray spectrum exposure time and time interval, lower limit on the intrinsic X-ray absorbing column density assuming $z=0$. We used PC mode spectra in the specified time interval, except for GRB~051227, GRB~080503, GRB~090515 and GRB~090607 where we used WT mode spectra. GRB~051227 and GRB~080503 have an extended emission ($^\mathrm{EE}$). For GRB~101219A we can not determine if it is SL or LL, due to the lack of the X-ray flux information at $10^4\,\mathrm{s}$ after the trigger.}
\label{tab:sample_lowerlimits}
\end{table}

\par Further confirmation that SGRBs are intrinsically absorbed is given by examining the Sample \MakeUppercase{\romannumeral 2}. We noticed that even if we put these SGRBs to $z=0$ and thus obtain the lower limits on the X-ray absorption, we get significant absorption in excess of the Galactic value (Table \ref{tab:sample_lowerlimits}). These $11$ lower limits are marked as black arrows in Figure \ref{fig:nhdist}. 

\par Having derived the intrinsic X-ray absorbing column densities, we checked if there is any dependence between $\mathrm{N_H}(z)$ and $\mathrm{T}_{90}$ and between $\mathrm{N_H}(z)$ and redshift, but there is no significant correlation. The Spearman's correlation coefficients are $\rho _\mathrm{T_{90}} ^{\mathrm{N_H}(z)} =-0.11 \pm 0.36$ ($p=0.80$) and $\rho _{z} ^{\mathrm{N_H}(z)} =0.14 \pm 0.35$ ($p=0.73$), respectively. The uncertainties on the correlation coefficients are quite large, but this is expected given the large uncertainties on the values of $\mathrm{N_H}(z)$.

\par We checked if $\mathrm{N_H}(z)$ correlates with the Galactic $\mathrm{N_H}$ or with photon index ($\Gamma$). We found no significant correlation, with the Spearman's correlation coefficients being $\rho _\mathrm{N_H(Gal.)} ^{\mathrm{N_H}(z)} =-0.47 \pm 0.32$ ($p=0.24$) and $\rho _\Gamma ^{\mathrm{N_H}(z)} =-0.52 \pm 0.36$ ($p=0.18$), respectively.

\subsection{Intrinsic X-ray absorption and host galaxy offset}
\par
\begin{figure*}
\begin{center}
\includegraphics[angle=-90,width=0.49\textwidth]{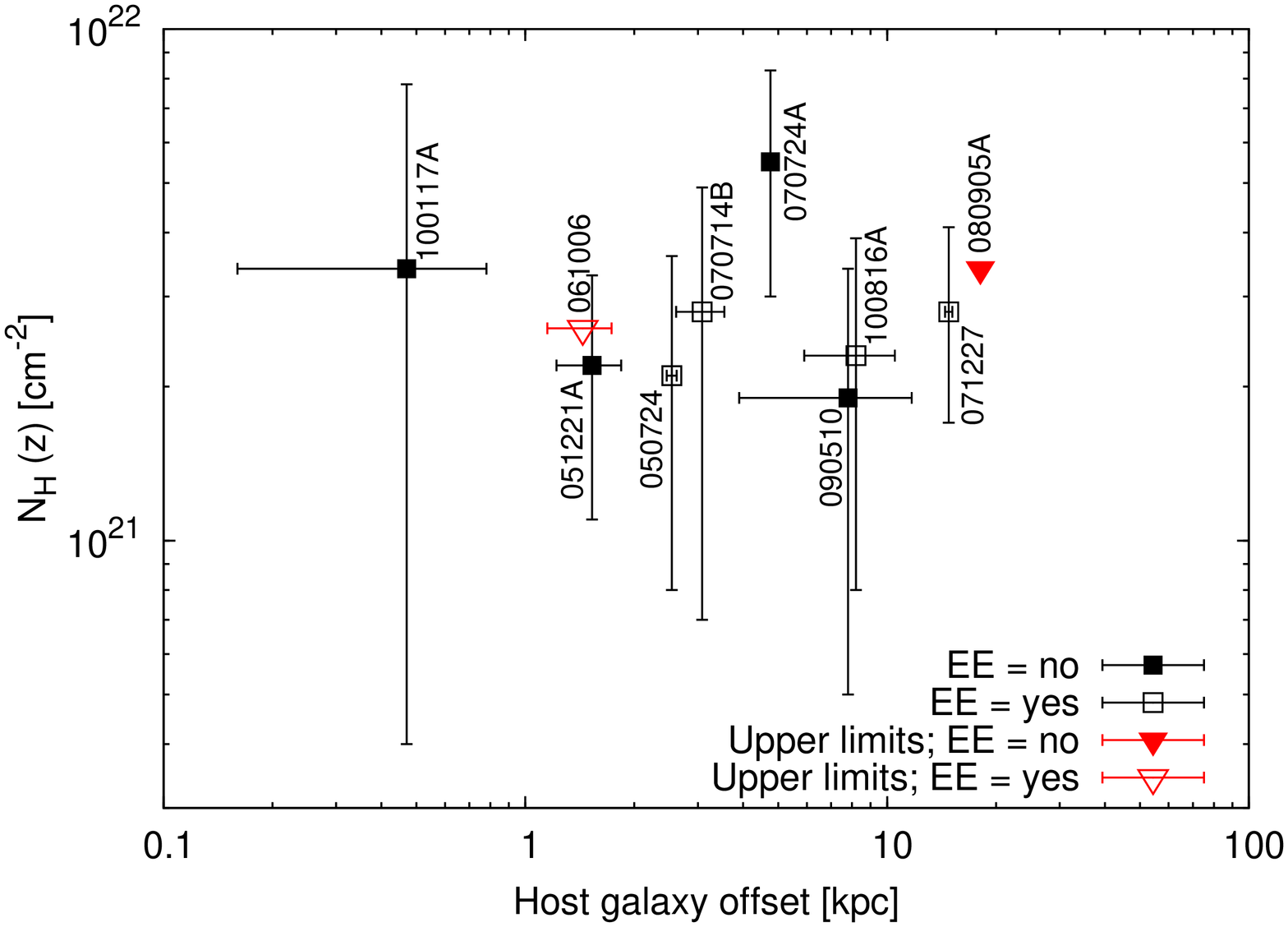}
\includegraphics[angle=-90,width=0.49\textwidth]{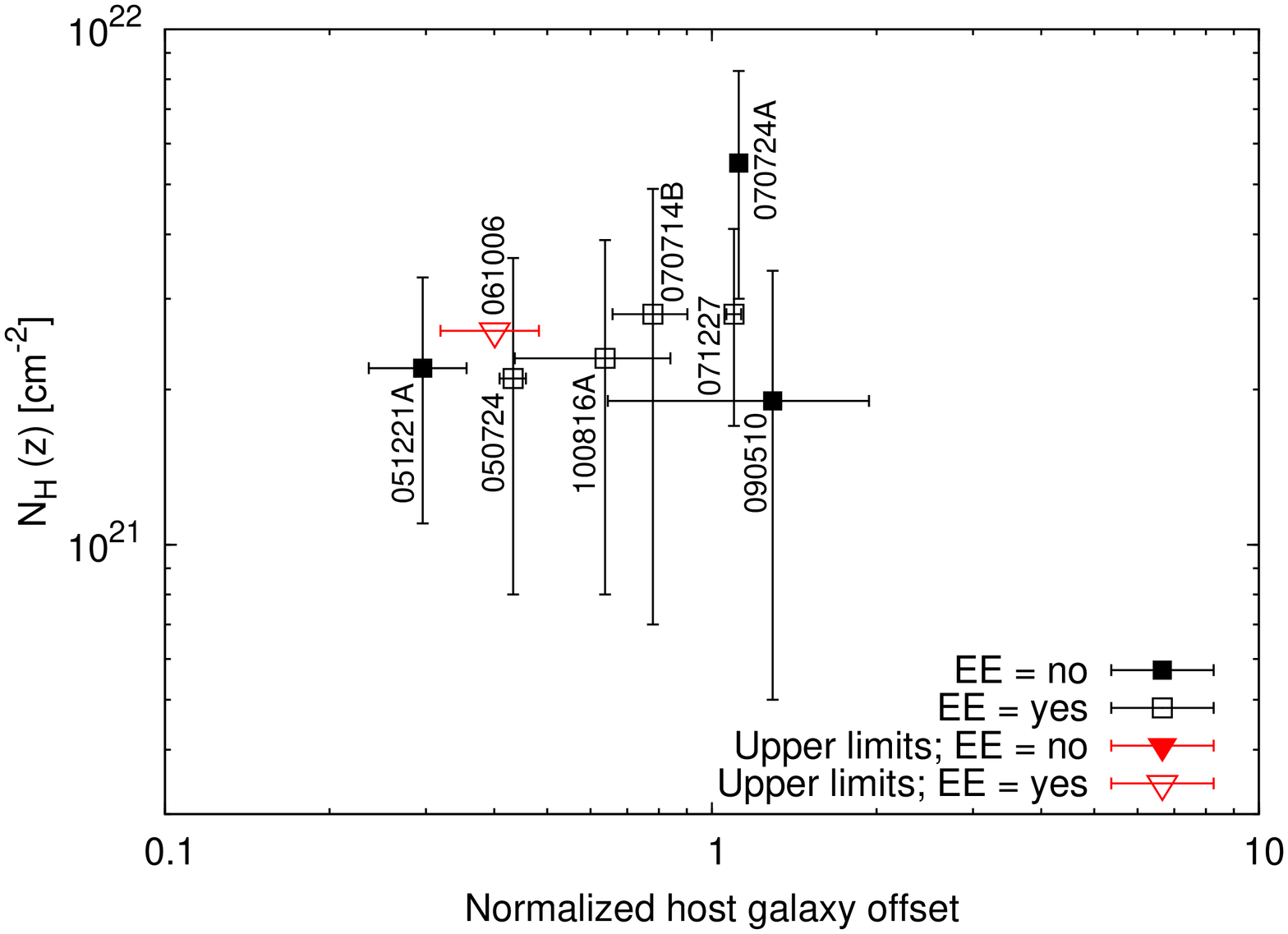}
\end{center}
\caption{Left: Intrinsic X-ray absorption versus host galaxy offset. Right: intrinsic X-ray absorption versus normalised host galaxy offset. Values are taken from Table \ref{tab:sample}. Filled points are SGRBs without an extended emission (EE), while empty points are those with an extended emission. Red points represent upper limits on $\mathrm{N_H}(z)$.}
\label{fig:nhvsoffset}
\end{figure*}

 For SGRBs from Sample \MakeUppercase{\romannumeral 1} we can investigate if there is any correlation between $\mathrm{N_H}(z)$ and host galaxy offset. Figure \ref{fig:nhvsoffset} shows $\mathrm{N_H}(z)$ versus host galaxy offsets and $\mathrm{N_H}(z)$ versus normalised host galaxy offsets.

\par We tested the possible correlations between (normalised) host galaxy offsets and $\mathrm{N_H}(z)$ using the Spearman's rank correlation test for Sample \MakeUppercase{\romannumeral 1}, excluding the $2$ SGRBs that have only upper limits on $\mathrm{N_H}(z)$. For host galaxy offsets versus $\mathrm{N_H}(z)$, the Spearman's correlation coefficient is $\rho _\mathrm{offsets} ^{\mathrm{N_H}(z)} = -0.07 \pm 0.36$ ($p=0.87$). For normalised host galaxy offsets versus $\mathrm{N_H}(z)$, the Spearman's correlation coefficient is $\rho _\mathrm{norm. \, offsets} ^{\mathrm{N_H}(z)} = 0.20 \pm 0.35$ ($p=0.67$). If we exclude GRB~070724A and GRB~071227, because the lower limits on their normalised host galaxy offsets are larger than $1$ (this would indicate that a GRB occurs outside of its host galaxy, if characterised by the effective radius), the Spearman's correlation coefficient is $\rho _\mathrm{norm. \, offsets} ^{\mathrm{N_H }(z)} = -0.10 \pm 0.47$ ($p=0.95$). All values indicate that there is no significant correlation. The uncertainties on the correlation coefficients are large due to large uncertainties on the values of $\mathrm{N_H}(z)$ and (normalised) host galaxy offsets (see Figure \ref{fig:nhvsoffset}).

\par For normalised host galaxy offsets, which indeed better represent the relative location of a GRB inside its host galaxy, one would expect that (especially when smaller than $1$) they will be inversely correlated with $\mathrm{N_H}(z)$. This would indicate that more we go towards the edge of the galaxy, less intrinsic absorption we have, i.e., the environment gets less dense. We obtained no such result. The reason could be either the large errors or that here we are dealing with projected offsets, which do not give us any information about the position of a GRB in its host galaxy along the line of sight. For the same projected offset, a GRB can occur closer or further away from us in its host galaxy, resulting in less or more intervening host galaxy material along the line of sight. Such effect is of course less strong near the edges of the galaxies, as is likely the case of GRB~071227, for which an optical afterglow position falls at the very edge of the galaxy disk \citep{DAvanzo2009}.

\par Furthermore, we note that the above analysis is probably affected by some selection effect against large offsets (and likely lower $\mathrm{N_H}(z)$) because, in order to have events with robust redshift, SGRBs of Sample \MakeUppercase{\romannumeral 1} include only events with positional coincidence between the optical afterglow and the host galaxy light (see Section \ref{sect:sample_selection}).

\subsection{X-ray afterglow brightness and the properties in gamma regime}
\label{sect:heprop}
\par As pointed out in Sections \ref{sect:sample_selection} and \ref{sect:intrinsic_absorption}, $44$ SGRBs promptly observed by the {\it Swift} XRT can be divided into two classes, according to the brightness of their X-ray afterglow. Almost half of them have X-ray afterglow bright enough to perform X-ray spectroscopy (they have at least $200$ counts with constant $0.3-1.5\,\mathrm{keV}$/$1.5-10\,\mathrm{keV}$ hardness ratio in the specified time interval), while the other half have too faint X-ray afterglow, and thus no spectroscopic study could be performed. Moreover, among the ``faint'' SGRBs, six events (i.e., $14\%$ of the whole sample or $26\%$ of the ``faint'' SGRBs' sample) have no X-ray afterglow detected, in spite of the prompt {\it Swift} XRT follow-up. This is at variance with respect to the LGRBs, where for $\sim 95\%$ of the events promptly observed by the {\it Swift} XRT an X-ray afterglow is detected \citep{Evans2009}. Such a bimodality in the X-ray afterglow brightness can be explained by differences in the GRB energetics (with more energetic GRBs having brighter afterglows) or in the density of the circumburst medium (with GRB exploding in denser environments having brighter afterglows).

\par In order to check if these two classes of SGRBs show any significant difference in their prompt emission properties, we computed their distribution of fluence, 1-s peak photon flux and $\mathrm{T}_{90}$, measured by the {\it Swift} BAT, and compared them with a two-sample Kolmogorov-Smirnov (K-S) test. As can be seen in Table \ref{tab:he_properties}, SGRBs with brighter X-ray afterglows seem to have, on average, higher fluences, peak fluxes and longer durations. The probabilities associated with the K-S tests for each distribution is of the order of a few percent (see Table \ref{tab:kstest2}, upper part), likely suggesting that the two classes of SGRBs have different prompt emission properties. However, the presence of $8$ SGRBs with an extended emission (GRB~050724, GRB~051227, GRB~061006, GRB~070714B, GRB~071227, GRB~080123, GRB~080503 and GRB~100816A) introduce some biases in this study. In particular, SGRBs with an extended emission can spuriously increase the average value of fluence and $\mathrm{T}_{90}$. We thus repeated the K-S tests excluding the events showing an extended emission. We find that the associated probabilities increase, suggesting that the two classes of SGRBs do not show significant differences in their prompt emission properties (see Table \ref{tab:kstest2}, lower part). 

\par The different brightness of their X-ray afterglows might thus be a consequence of the density of the environment around the GRB and being indicative of different progenitors. A possible theoretical explanation would be that we are dealing with two distinct NS-NS or NS-BH populations. The events showing brighter X-ray afterglows might be associated with primordial double compact object systems merging in a relatively short time (and thus occurring inside their host galaxies, in star forming environments; \citealt{PernaBelczynski2002}), while the events with fainter X-ray afterglows might be originated by double compact object systems which experienced a large natal kick or which are dynamically formed in globular clusters (associated with a low-density environments; see Section \ref{sect:intro}). 

\par On the other hand, we note that the values of $\mathrm{T}_{90}$, fluence and 1-sec peak photon flux reported in Table \ref{tab:he_properties} are measured in the observer's frame. It has not been possible, due to the lack of redshift measurement for many SGRBs, to transform these values to the GRB's rest frame, especially for SGRBs with faint X-ray afterglow. This may again introduce some bias, especially for SGRBs that would happen at high redshift \citep{Berger2007}. 

\begin{table}
\begin{center}
\tabcolsep 24pt
\begin{tabular}{ccc}
\hline
Distribution & $D$ & $P$ \\
\hline
\multicolumn{3}{c}{All SGRBs:} \\
$\mathrm{T}_{90}$ & $0.43$ & $0.02$ \\
Fluence & $0.43$ & $0.02$ \\
1-sec peak flux & $0.31$ & $0.20$ \\
& & \\
\multicolumn{3}{c}{SGRBs without an extended emission:} \\
$\mathrm{T}_{90}$ & $0.40$ & $0.09$ \\
Fluence & $0.30$ & $0.34$ \\
1-sec peak flux & $0.30$ & $0.36$ \\
\hline
\end{tabular}
\end{center}
\caption{Results of the two-sample Kolmogorov-Smirnov test from comparing $\mathrm{T}_{90}$, fluence and 1-sec peak photon flux distributions between ``bright'' (Sample \MakeUppercase{\romannumeral 1} and \MakeUppercase{\romannumeral 2}) and ``faint'' X-ray afterglow samples. Upper part are results for all SGRBs, while lower part are results for SGRB without an extended emission (see Table \ref{tab:he_properties}). $D$ represents the calculated value of the Kolmogorov-Smirnov statistic, and $P$ is the corresponding probability that two distributions are not drawn from different populations.} 
\label{tab:kstest2}
\end{table}
 
\subsection{SGRBs with an extended emission}

\par
\begin{figure}
\begin{center}
\includegraphics[angle=-90,width=1\columnwidth]{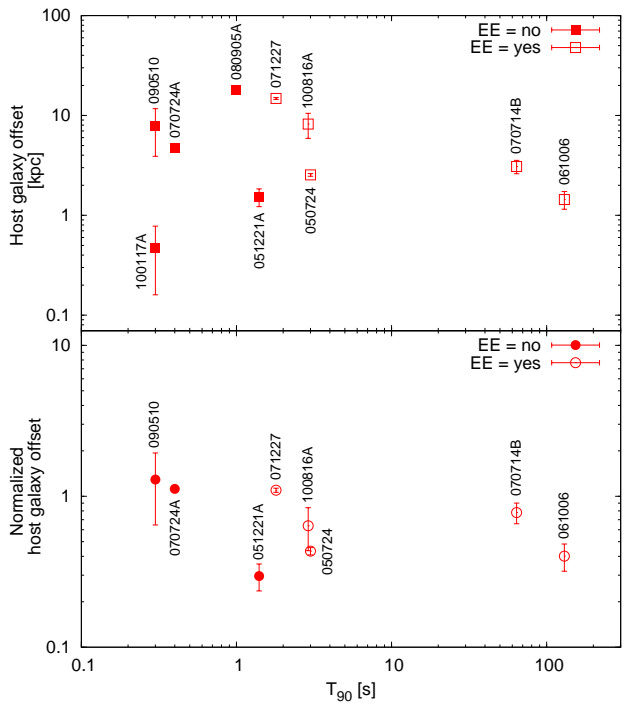}
\end{center}
\caption{Host galaxy offset (top part, squares) and normalised host galaxy offset (bottom part, circles) versus $\mathrm{T}_{90}$ for SGRBs from Sample \MakeUppercase{\romannumeral 1}. Filled symbols represent SGRBs without an EE, while empty symbols represent SGRBs with an EE.}
\label{fig:normhostvst90}
\end{figure}

 To investigate if SGRBs with an EE from Sample \MakeUppercase{\romannumeral 1} lie on average closer to the centre of their host galaxies (as presented first by \citealt{Troja2008}), we plotted in Figure \ref{fig:normhostvst90} the $\mathrm{T}_{90}$ values versus host galaxy and normalised host galaxy offsets. The value of $\mathrm{T}_{90}$ somehow indicates the duration of a GRB, and SGRBs with an EE have on average larger $\mathrm{T}_{90}$ than SGRBs without an EE. Calculating the Spearman's rank correlation coefficient between $\mathrm{T}_{90}$ and host galaxy offsets, we obtain $\rho _\mathrm{offsets} ^\mathrm{EE} = -0.13 \pm 0.10$ ($p=0.73$). This shows that there is no clear anti-correlation between $\mathrm{T}_{90}$ and host galaxy offset. Similar conclusions were also presented by \citet{Fong2010}.

\par Extending this analysis, we checked if there is any correlation between $\mathrm{T}_{90}$ and normalised host galaxy offset (see Figure \ref{fig:normhostvst90}, bottom panel). In this case the data look somewhat less scattered. Calculating the Spearman's rank correlation coefficient (without GRB~100117A), we obtain $\rho _\mathrm{norm. \, offsets} ^\mathrm{EE} = -0.57 \pm 0.20$ ($p=0.15$). For normalised host galaxy offsets, there is a hint of anti-correlation with $\mathrm{T}_{90}$.

\par Physical explanation for such correlation is not entirely known at present. \citet{Troja2008} suggested that the anti-correlation between $\mathrm{T}_{90}$ and host galaxy offset may be due to different progenitors of SGRBs, arguing that SGRBs without an EE occur via NS-NS mergers. These have larger offsets from their host galaxy centre, while SGRBs with an EE occur via NS-BH mergers. Opposite to that, \citet{Church2011} argued that both BH-NS and NS-NS mergers have similar offset distributions, and \citet{PernaBelczynski2002} argued that different groups of NS-NS mergers exist, with a significant fraction of them merging well within their host galaxies. \citet{Norris2010} proposed that an EE, which is observed in $\sim 25\%$ of SGRBs discovered by the \textit{Swift} satellite, is a part of the prompt emission, probably not directly caused by the properties of the surrounding environment. Their explanation for SGRBs' dichotomy concerning an EE might lie in the physical properties of a compact binary merger (e.g., mass, angular momentum, etc.), while the progenitor type is only one. Alternatively, the temporally long-lasting soft tail observed in SGRBs could be originated by the afterglow emission and related to the density of the circumburst medium \citep{Bernardini2007, Caito2009, Caito2010, deBarros2011}.

\subsection{Short-lived/Long-lived X-ray Afterglow}
\label{sl_ll}
\par \citet{Sakamoto2009} found that there exist two distinct classes of SGRBs based on the duration of the X-ray afterglow, namely short-lived (SL), for which the X-ray afterglow flux at $10^4\,\mathrm{s}$ after the trigger is less than $10^{-13}\,\mathrm{erg\,cm^{-2}\,s^{-1}}$, and long-lived (LL), for which the X-ray afterglow flux is more than $10^{-13}\,\mathrm{erg\,cm^{-2}\,s^{-1}}$. Based on the sample therein, \citet{Sakamoto2009} concluded that SL SGRBs show no extended emission, no optical afterglow and a large host galaxy offset.
\par We checked if this classification holds for SGRBs in our samples. Among them, there is GRB~100117A (Sample \MakeUppercase{\romannumeral 1}), which based on the X-ray light curve and according to the definition of \citet{Sakamoto2009} is SL, but it has an OA detected and a very small host galaxy offset. Besides that, in Sample \MakeUppercase{\romannumeral 2} there are GRB~090515 (\citealt{Rowlinson2010}), which is SL and shows no EE, but has an OA detected, and GRB~080503 (\citealt{Perley2009}), which is SL, but shows an EE and has an OA detected. Another possible candidate could be GRB~080905A, which has an OA detected, but we can not confirm if it is SL due to the lack of the X-ray flux information around $10^4\,\mathrm{s}$ after the trigger, although its light curve looks very similar to the light curves of other SL SGRBs.
\par We therefore conclude that there are short-lived SGRBs (GRB~080503, GRB~090515 and GRB~100117A), which have an OA detected and some also have an EE. According to these findings, the X-ray afterglow duration does not seem to be an unique indicator of a specific progenitor and/or an environment for SGRBs.

\section{Conclusion}
\label{sect:conclusion}
\par We presented here a systematic study of the environment of SGRBs performed through the measure of their intrinsic X-ray absorbing column densities. Our results show that there are possibly two distinct populations of SGRBs: half of them (Sample \MakeUppercase{\romannumeral 1} with robust redshifts and Sample \MakeUppercase{\romannumeral 2} without robust redshifts) show relatively bright X-ray afterglows and occur in dense environments, with X-ray column densities comparable to the one measured for LGRBs in the same redshift range. Another half of them have very faint or no X-ray afterglow at all. The properties in the gamma regime of these two classes of SGRBs do not differ significantly, possibly suggesting that the observed difference in their X-ray afterglow brightness is not due to the burst energetics and might be a consequence of the environment where they explode.

\par We found no correlation between the intrinsic X-ray absorption and the host galaxy offset or normalised host galaxy offset. This is perhaps not expected for normalised host galaxy offsets, however the results could be affected by the large uncertainties on the values of $\mathrm{N_H}(z)$ or by the fact that we are dealing with the projected offsets, which do not give us absolute position of a GRB inside its host galaxy along the line of sight.

\par We checked if SGRBs with an extended emission lie closer to the centre of their host galaxies. We found that there is no significant anti-correlation between duration ($\mathrm{T}_{90}$) and host galaxy offset, but that there is a hint of anti-correlation when using normalised host galaxy offsets instead. 

\par We also tested if short-lived SGRBs from our samples have no extended emission, no optical afterglow and large host galaxy offset, as proposed by \citet{Sakamoto2009}. We found that this does not hold for all cases, since GRB~080503, GRB~090515 and GRB~100117A are SL, but have an OA detected. GRB~080503 has also an EE, while GRB~100117A has a very small host galaxy offset.

\par Finally, we want to stress that the sample of SGRBs with a robust redshift determination and host galaxy offset measured is quite small up to this date (Sample \MakeUppercase{\romannumeral 1} is about nine times smaller than the whole sample of LGRBs with known redshift from \citet{Campana2010}, and almost twice as small as the subsample of LGRBs with $z<0.923$), mainly due to their dim optical afterglows. Thus, the results might be affected by the size of the sample. Future data will show if the conclusions, drawn in this paper, also hold on a larger sample.

\section*{Acknowledgments}
This work made use of data supplied by the UK Swift Science Data Centre at the University of Leicester. We thank J.F. Graham for providing Gemini-N images of GRB 070714B. DK is grateful to the staff at INAF-OAB in Merate for their hospitality during his four-month visit, and acknowledges funding from the Slovene human resources development and scholarship fund (grant 11012-10/2011). AG acknowledges funding from the Slovenian Research Agency and from the Centre of Excellence for Space Sciences and Technologies SPACE-SI, an operation partly financed by the European Union, European Regional Development Fund and Republic of Slovenia, Ministry of Higher Education, Science and Technology. PDA, AM, SCampana, SCovino, MGB, SDV, RS and GT acknowledge the Italian Space Agency for
financial support through the project ASI I/004/11/0

\begin{table*}
\begin{center}
\begin{tabular}{cccccccc}
\hline
GRB & 
$\mathrm{T}_{90}$ &
Fluence & 
1-s peak photon flux & 
GRB & 
$\mathrm{T}_{90}$ &
Fluence & 
1-s peak photon flux
\\
& 
[s] &  
[$10^{-7}\,\mathrm{erg/cm^2}$] &
[$\mathrm{ph/cm^2/sec}$] &
& 
[s] &  
[$10^{-7}\,\mathrm{erg/cm^2}$] &
[$\mathrm{ph/cm^2/sec}$]
\\
\hline
\multicolumn{4}{c}{``Bright'' X-ray afterglow:} & \multicolumn{4}{c}{``Faint'' X-ray afterglow:}\\
050724$^\mathrm{EE}$ & $3.0$ & $9.98 \pm 1.20$ & $3.26 \pm 0.30$ & 050509B & $0.048$ & $0.09 \pm 0.02$ & $0.28 \pm 0.10$ \\
051221A	& $1.4$ & $11.50 \pm 0.35$ & $12.00 \pm 0.39$ & 050813 & $0.6$ & $0.44 \pm 0.11$ & $0.94 \pm 0.23$ \\
051227$^\mathrm{EE}$ & $8.0$ & $6.99 \pm 1.08$ & $0.95 \pm 0.12$ & 050906 & $0.128$ & $0.06 \pm 0.02$ & $0.22 \pm 0.11$ \\
060313 & $0.7$ & $11.30 \pm 0.45$ & $12.10 \pm 0.45$ & 050925 & $0.068$ & $0.76 \pm 0.09$ & $10.00 \pm 1.19$ \\
060801 & $0.5$ & $0.80 \pm 0.10$ & $1.27 \pm 0.16$ & 051105A & $0.028$ & $0.22 \pm 0.04$ & $0.32 \pm 0.12$ \\
061006$^\mathrm{EE}$ & $130.0$ & $14.20 \pm 1.42$ & $5.24 \pm 0.21$ & 051210 & $1.3$ & $0.85 \pm 0.14$ & $0.75 \pm 0.12$ \\
070714B$^\mathrm{EE}$ & $64.0$ & $7.20 \pm 0.90$ & $2.70 \pm 0.20$ & 060502B & $0.09$ & $0.40 \pm 0.05$ & $0.62 \pm 0.12$ \\
070724A	& $0.4$ & $0.30 \pm 0.07$ & $1.00 \pm 0.20$ & 061201 & $0.8$ & $3.34 \pm 0.27$ & $3.86 \pm 0.31$ \\
071227$^\mathrm{EE}$ & $1.8$ & $2.20 \pm 0.30$ & $1.60 \pm 0.20$ & 061217 & $0.3$ & $0.42 \pm 0.07$ & $1.49 \pm 0.24$ \\
080503$^\mathrm{EE}$ & $170.0$ & $20.00 \pm 1.00$ & $0.90 \pm 0.10$ & 070209 & $0.1$ & $0.22 \pm 0.05$ & $0.38 \pm 0.13$ \\
080702A & $0.5$ & $0.36 \pm 0.10$ & $0.70 \pm 0.20$ & 070729 & $0.9$ & $1.00 \pm 0.20$ & $1.00 \pm 0.20$ \\
080905A & $1.0$ & $1.40 \pm 0.20$ & $1.30 \pm 0.20$ & 070809 & $1.3$ & $1.00 \pm 0.10$ & $1.20 \pm 0.20$ \\
080919 & $0.6$ & $0.72 \pm 0.11$ & $1.20 \pm 0.20$ & 070810B & $0.08$ & $0.12 \pm 0.03$ & $1.80 \pm 0.40$ \\
081226A & $0.4$ & $0.99 \pm 0.18$ & $2.40 \pm 0.40$ & 080123$^\mathrm{EE}$ & $115.0$ & $5.70 \pm 1.70$ & $1.80 \pm 0.40$ \\
090510 & $0.3$ & $3.40 \pm 0.40$ & $9.70 \pm 1.10$ & 081024A & $1.8$ & $1.20 \pm 0.20$ & $1.10 \pm 0.10$\\
090515 & $0.036$ & $0.20 \pm 0.03$ & $5.70 \pm 0.90$ & 090305A & $0.4$ & $0.75 \pm 0.13$ & $1.90 \pm 0.40$ \\
090607 & $2.3$ & $1.10 \pm 0.20$ & $0.70 \pm 0.10$ & 090621B & $0.14$ & $0.70 \pm 0.10$ & $3.90 \pm 0.50$ \\
100117A	& $0.3$ & $0.93 \pm 0.13$ & $2.90 \pm 0.40$ & 091109B & $0.27$ & $1.90 \pm 0.20$ & $5.40 \pm 0.40$ \\
100702A & $0.16$ & $1.20 \pm 0.10$ & $2.00 \pm 0.20$ & 100206A & $0.12$ & $1.40 \pm 0.20$ & $1.40 \pm 0.20$ \\
100816A$^\mathrm{EE}$ & $2.9$ & $20.00 \pm 1.00$ & $10.90 \pm 0.40$ & 100213A & $2.4$ & $2.70 \pm 0.30$ & $2.10 \pm 0.20$ \\
101219A & $0.6$ & $4.60 \pm 0.30$ & $4.10 \pm 0.20$ & 100625A & $0.33$ & $2.30 \pm 0.20$ & $2.60 \pm 0.20$ \\
 & & & & 100628A & $0.036$ & $0.25 \pm 0.05$ & $0.50 \pm 0.10$ \\
  & & & & 101224A & $0.2$ & $0.58 \pm 0.11$ & $0.70 \pm 0.20$ \\
\hline
\end{tabular}
\end{center}
\caption{BAT $\mathrm{T}_{90}$ (in the $15-350\,\mathrm{keV}$ energy band), fluence and 1-sec peak photon flux (from the time-averaged spectrum, in the $15-150\,\mathrm{keV}$ energy band) for $44$ SGRBs that were promptly observed by the \textit{Swift} XRT up to January 2011. Left column represents SGRBs from Sample \MakeUppercase{\romannumeral 1} and \MakeUppercase{\romannumeral 2}, while right column represents all other SGRBs. $^\mathrm{EE}$ indicates SGRBs with an extended emission. Errors at $90\%$ confidence level are given for fluence and 1-sec peak photon flux. The table excludes $3$ GRBs for which the classification as a SGRB is debated: GRB~060614, GRB~090426 and GRB~100724A, as well as $13$ SGRBs which were not observed promptly by the \textit{Swift} XRT: GRB~050709, GRB~051114, GRB~061210, GRB~070406, GRB~070429B, GRB~071112B, GRB~080121, GRB~081101, GRB~090531B, GRB~090815C, GRB~090916, GRB~091117 and GRB~100216A}
\label{tab:he_properties}
\end{table*}

\end{document}